\documentclass{article}

\usepackage{PRIMEarxiv}

\usepackage[utf8]{inputenc} 
\usepackage[T1]{fontenc}    
\usepackage{url}            
\usepackage{booktabs}       
\usepackage{amsfonts}       
\usepackage{nicefrac}       
\usepackage{microtype}      
\usepackage{fancyhdr}       
\usepackage{graphicx}       
\usepackage{amsmath,amssymb,amsthm}
\usepackage{multirow}
\usepackage{fontawesome5}
\usepackage{algorithm}
\usepackage{algpseudocode}
\usepackage[
  colorlinks=true,
  linkcolor=blue,
  citecolor=blue,
  urlcolor=blue
]{hyperref}
\usepackage{doi}
\usepackage[
  style=authoryear, 
  backend=biber, 
  sorting=nyt, 
  natbib=true,
  doi=true,
  isbn=false,
  eprint=false,
  minnames=1, 
  maxnames=2,
  url=false      
]{biblatex}

\AtEveryBibitem{\clearfield{issn}\clearfield{urldate}}
\title{NONPARAMETRIC SAFETY STOCK DIMENSIONING: A DATA-DRIVEN APPROACH FOR SUPPLY CHAINS OF HARDWARE OEMS
}

\author{
  Elvis Agbenyega\thanks{Lenovo. ISG GSC SmartOps. Corresponding author.} \\ 
    eagbenyega@lenovo.com \\
  Lenovo Infrastructure Solutions Group \\
  Morrisville, NC, USA\\
  \and 
  Cody Quick\thanks{Lenovo. ISG GSC SmartOps.} \\
      cquick@lenovo.com \\
  Lenovo Infrastructure Solutions Group \\
  Morrisville, NC, USA\\
}

\begin{document}
\maketitle

\begin{abstract}
    Resilient supply chains are critical, especially for Original Equipment Manufacturers (OEMs) that power today's digital economy. Safety Stock dimensioning---the computation of the appropriate safety stock quantity---is one of several mechanisms to ensure supply chain resiliency, as it protects the supply chain against demand and supply uncertainties. Unfortunately, the major approaches to dimensioning safety stock heavily assume that demand is normally distributed and ignore future demand variability, limiting their applicability in manufacturing contexts where demand is non-normal, intermittent, and highly skewed. In this paper, we propose a data-driven approach that relaxes the assumption of normality, enabling the demand distribution of each inventory item to be analytically determined using Kernel Density Estimation. Also, we extended the analysis from historical demand variability to forecasted demand variability. We evaluated the proposed approach against a normal distribution model in a near-world inventory replenishment simulation. Afterwards, we used a linear optimization model to determine the optimal safety stock configuration. The results from the simulation and linear optimization models showed that the data-driven approach outperformed traditional approaches. In particular, the data-driven approach achieved the desired service levels at lower safety stock levels than the conventional approaches.. \\

\end{abstract}


\faGithub\ \textbf{Code:} \url{https://github.com/graphshade/safety_stock_simulation} \\ \\

\section{Introduction}\label{sec:intro}

Supply chain resiliency is crucial for Original Equipment Manufacturers (OEMs), especially those that produce essential technological products powering today's digital economy and other critical infrastructure sectors, such as financial services, transportation systems, the defense industrial base, and energy. One inventory management mechanism used by OEMs to ensure supply chain resiliency is safety stock dimensioning—the setting of optimal safety stock quantity to buffer against demand and supply uncertainty. Setting the correct safety stock quantity guarantees high supply chain serviceability and reduced inventory holding costs. 

Existing approaches for dimensioning safety stock can be classified into four categories: dimensioning safety stock using demand variance, dimensioning safety stock using forecast error variance, dimensioning safety stock using product organization and standardization, and dimensioning safety stock using mathematical programming and neural networks~\citep{goncalves_operations_2020}. These approaches are not without limitations. For instance, dimensioning safety stock using demand variance —the predominant approach in the empirical literature —assumes normality of demand patterns. In the real world, demand is non-normal, intermittent, and highly skewed~\citep{barros_systematic_2021}. Similarly, modelling safety stock with forecast-error variance is susceptible to bias when the forecast errors are autocorrelated. The other approaches---using mathematical programming and product organization---are limited when the underlying demand distribution deviates from normality. Also, existing approaches ignore future demand variance, which, in practice, contradicts the underlying notion of safety stock---a buffer stock that protects the supply chain against future demand variability.  

The limitations of existing approaches to dimensioning safety stock are profound in manufacturing contexts where component demand is non-normal, intermittent, or skewed. In such contexts, assuming demand is normally distributed and ignoring future demand variability results in safety stock settings that either overestimate or underestimate safety stock requirements, causing the supply chain to deviate from desired service-level agreements. 

In this paper, we propose a data-driven approach leveraging Kernel Density Estimation (KDE). This non-parametric density estimation method makes no assumptions about the distribution of per-component demand. The estimated densities are converted to probability mass function using procedures outlined by
\citet{aitchison_multivariate_1976}, \citet{wand_kernel_1994}, and \citet{chen_probability_2000}. We input the attributes of the estimated probability mass function into a simulation model to evaluate the performance of the proposed model against a normal model and the classical approach to computing safety stock under approximate inventory replenishment cycle scenarios. A linear optimization model was later used to determine the optimal safety stock configuration under both the proposed data-driven approach and the normal model. 

The remainder of the paper is structured as follows. In Section~\ref{sec:related_work}, we examined related works critically, identifying various gaps in extant approaches to modelling safety stock. In Section~\ref{sec:methods}, we expounded on the proposed data-driven approach methodically. After that, we applied the proposed methodology to a sample dataset from a server hardware OEM and reported the results in Section~\ref{sec:results}. Section~\ref{sec:discussion} discussed the managerial implications of the results and offers insight into the proposed approach. Finally, in Section~\ref{sec:conclusion}, we summarize the paper and provide directions for future research and empirical analysis.

\section{Related Work}\label{sec:related_work}

\subsection{Safety Stock Dimensioning}\label{subsec:rw_ss_dim}

Supply chains constantly face risks from demand and supply uncertainties~\citep{demiray_kirmizi_enhancing_2024, fan_change_2025}. One primary mechanism used by supply chains to manage the risk from demand and supply uncertainties is inventory management through safety stock dimensioning, \citet{koh_uncertainty_2002} explained safety stock dimensioning to mean the setting of the optimal safety stock quantity for individual items. Throughout the extant literature, significant work has been done to address the problem of safety stock dimensioning. \citet{goncalves_operations_2020} identified four categories of strategies for calculating safety stock: dimensioning based on demand variation, dimensioning based on forecast error variability, dimensioning based on product structure and standardization, and dimensioning based on mathematical programming, simulation, and neural networks. In the following sections, we critically examine each dimensioning approach, assessing its strengths and weaknesses and identifying gaps in methodology that this paper proposes to address.

\subsection{Dimensioning with Variation in Demand}\label{subsec:rw_demand_var}

The classical formulation of safety stock dimensioning, as stated in (\ref{eq:eq_trad_1}) (see \citet{alicke_planung_2005} for a detailed description), is premised on calculating a quantity of buffer stock that enables the supply chain to meet a targeted service level ($\alpha$) on a product. To do this, a service level factor, z-score ($z_\alpha$), is determined from the standard normal distribution as the number of standard deviations above the mean demand corresponding to the targeted serviceability. For example, a targeted service level of 95\% is equal to a service level factor of 1.65. The service level factor is then multiplied by the demand variation, as measured by the standard deviation of historical demand.

\begin{equation}
    S_s = z_\alpha \cdot \sigma_D
    \label{eq:eq_trad_1}
\end{equation}
where $S_s$ = safety stock level (units), $\alpha$ = service level, $z_\alpha$ = safety factor depending on $\alpha$, $\sigma_D$ = standard deviation of demand (units/period). Method~(\ref{eq:eq_trad_1}) was extended to include the total replenishment time, resulting in (\ref{eq:eq_trad_2})~\citep{alicke_planung_2005}. 

\begin{equation}
    S_s = z_\alpha \cdot \sigma_D \cdot \sqrt{L+R}
    \label{eq:eq_trad_2}
\end{equation}
where $L$ = lead time (in calendar period) and $R$ = replenishment review period (in calendar period).

While the classical formulation, as noted in (\ref{eq:eq_trad_1}) and (\ref{eq:eq_trad_2}), is intuitive and straightforward, it makes several simplifying assumptions that limit its ability to model realistic scenarios~(\citep{eppen_determining_1988}. First, the assumption that demand is normally distributed rarely holds in most supply chains. Consequently, dimensioning safety stock under the assumption of normality often led to overstocking or understocking. As \citet{ruiz-torres_safety_2010} observed, incorrectly assuming normality results in higher inventory carrying costs. Second, assuming normal distribution does not extend to scenarios where demand is discrete rather than continuous or where demand cannot be negative~\citep{vandeput_inventory_2020}. For example, in the server manufacturing industry, the selected industry for this analysis, there is no such thing as decimal demand for a central processing unit (CPU) component or negative demand for a solid-state drive (SSD) component. Lastly, measuring demand variation using the standard deviation of historical demand is anathema to the fundamental notion of safety stock management, which is to determine and manage a buffer stock to meet future demand above the expected demand. Thus, forecast variance is ignored when determining the safety stock~\citep{derbel_empirical_2022,eppen_determining_1988, goncalves_multivariate_2021,prak_calculation_2017, trapero_empirical_2019}.

\subsection{Dimensioning with Forecast Error Variance}\label{subsec:rw_forecast_var}

One practical challenge in determining safety stock in the real world is estimating demand variability. Often, the actual demand variance is unknown. As a result, the traditional approach specified in (\ref{eq:eq_trad_1}) approximated demand variance using historical demand data, which is without no limitation. As noted by \citet{eppen_determining_1988}, using historical demand variance can lead to either underestimating or overestimating the safety stock estimate. To improve the estimation of demand variance, \citet{eppen_determining_1988} incorporated forecast error variance. Forecast error variance is measured as the standard deviation of the difference between demand and the forecast. The formulation of (\ref{eq:eq_trad_2}) using forecast error variance is restated in (\ref{eq:eq_trad_3}). The intuition for this mechanism is that forecast error is proportional to safety stock. Thus, the higher the forecast accuracy, the lower the level of safety stock needed to manage demand uncertainty.

\begin{equation}
    S_s = z_\alpha \cdot \sigma_F \cdot \sqrt{L+R}
    \label{eq:eq_trad_3}
\end{equation}
where $\sigma_F$ =  per period standard deviation of forecasted error for the demand over the total replenishment period (units/period).

\citet{eppen_determining_1988} assumed that forecast errors are normally distributed, which is rarely the practice case. \citet{ruiz-torres_safety_2010} used non-parametric kernel density estimates to relax the assumption of normality. Another challenge with using forecast errors is the presence of autocorrelation. When forecast errors are not independent and identically distributed, the standard deviation of the forecast errors becomes biased~\citep{silver_biased_1987}. According to \citet{prak_calculation_2017}, using forecast error variability without correcting for forecast error autocorrelation over the replenishment time period (lead time plus review period) leads to flawed safety stock outcomes. Likewise, \citet{ali_forecast_2012} reported that forecast variance significantly impacts safety stock dimension under conditions of intermittent demand.

\subsection{Dimensioning with Product Structure and Standardization}\label{subsec:rw_product_str}

Dimensioning of safety stock based on product structure and standardization is rarely studied~\citep{goncalves_operations_2020}. Product structure represents the hierarchical composition or arrangement of a product's sub-components. Through standardization, common product structures can affect safety stock requirements~\citep{collier_measurement_1981}. To illustrate the effect of component commonality on safety stock levels, \citet{collier_aggregate_1982} proposed an analytical method that measures commonality as the ratio of standard components to distinct components and compares these levels to simulated safety stock levels. In material requirements planning (MRP) systems, setting the appropriate safety stock is vital to reducing emergency runs. For instance, \citet{carlson_safety_1986} proposed heuristic algorithms for setting safety stock at components that most frequently trigger emergency production. \citet{hernandez-ruiz_optimizing_2016} extended extant literature on component commonality to propose a mathematical model based on group technology (GT) philosophy. GT philosophy organizes production by grouping parts with similar design or manufacturing characteristics into "families". The core idea is to lower inventory by capitalizing on similarities. The researchers modelled a bill of materials with three levels of complexity and included a GT substitution factor, expressed as an inverse function of the number of components (e.g., $S_{f} = 1 - 1/N_{c}$). The simulated scenarios showed a significant reduction in safety stock levels when the authors included GT philosophy alongside component commonality.

\subsection{Dimensioning based on Mathematical Programming, Simulation, and Neural Networks}\label{subsec:rw_others}

Other approaches to safety stock dimensioning include using mathematical programming, simulation and neural networks. The first attempt to use nonlinear programming to model safety stock was made by \citet{bourland_strategic_1994} in their seminal work on solving the stochastic economic lot scheduling problem. The nonlinear programming model included safety stock and capacity slack to minimize cost. \citet{bourland_strategic_1994} concluded that combining safety stock and idle time to manage demand uncertainty is inversely related to overtime. The major limitation of their approach is complexity. The study made simplifying assumptions about the distribution of production runs to make the mathematical model closed form. 

\citet{louly_calculating_2009} also used mathematical programming to model safety stock under random lead times in the context of assembly systems. Specifically, \citet{louly_calculating_2009} proposed using a branch-and-bound algorithm to determine the optimal safety stock for a component. While the study demonstrated the effectiveness of the branch-and-bound algorithm for discrete lead-time distribution, it was conducted using a single-level component assembly system. Also, \citet{louly_calculating_2009} assumed no finished goods inventory.
Simulation-based approaches are often leveraged to either optimize input parameters or analyze different scenarios for the mathematical programming approach~\citep{goncalves_operations_2020}. For instance, a seminar paper by \citet{avci_multi-objective_2017} proposed a decomposition-based multi-objective differential evolution algorithm (MODE/D) to model and compute safety stock levels that minimize holding costs and premium freight costs. When evaluated in the context of a global automotive company, the MODE/D approach yielded lower holding costs compared to traditional supply chain conditions. However, the applicability of  \citet{avci_multi-objective_2017} is limited to divergent supply chains with zero assembly production. \citet{avci_multi-objective_2018} extended the previous work on (MODE/D) to model convergent supply chains. The simulation methodological setup, first followed by optimization, was leveraged again. However, the approach in \citet{avci_multi-objective_2017} was extended to model dependent component demands characteristic of convergent supply chains. When evaluated in the context of a global automotive supply chain, the final holding cost and premium freight were lower than under traditional safety stock-diminishing strategies. 

Neural networks are increasingly used to model supply chain problems. For instance, neural networks are used in demand forecasting and supplier selection decision-making~\citep{kourentzes_neural_2014, kuo_integration_2010}. \citet{zhang_distributed_2017} applied neural networks to estimate safety stock levels in a warehouse product service system. The setup modelled five input features (selling frequency, storage cost, shortage cost, demand, purchasing quantity) that are likely to impact safety stock into a multi-layer perceptron to forecast safety stock level. The final model recorded an average error rate of 6.3\%. Since neural networks require a massive volume of data to train, this approach is not data-efficient. Also, if the safety stock levels used as targets in the training face are not optimal, the final model may likely learn suboptimal mappings~\citep{goncalves_operations_2020}.

\section{Methods}\label{sec:methods}

\subsection{Data Description}\label{subsec:mt_data_desc}

The data used for the analysis are in two parts: descriptive analytics (normality tests and hypothesis tests) and safety stock dimensioning. For descriptive analytics, the data comprises 52 weeks of historical consumption and corresponding forecasted demand over the same horizon. For safety stock dimensioning, the data include 52-week historical consumption, 13-week forward-looking forecasted demand, lead time in weeks, unit cost per item, and item class (class A for high-priority items, class B for second-level high-priority items, and class C for low-priority items). The data is collected on 20 purchase group subcomponents of an SSD category for a server assembly manufacturer.

\subsection{Normality Test of Item-Level Demand}\label{subsec:mt_normality_test}

The distribution of each item's demand is tested for normality to determine whether a parametric or nonparametric method is appropriate for safety stock estimation. The three major tests of normality---Shapiro-Wilk test, D'Agostino's $K^2$ test, and Anderson-Darling test---were applied. The decision criteria for the Shapiro-Wilk test and D'Agostino's K² test are to reject the null hypothesis that demand is usually distributed if the p-value is less than the 5\% level of significance. For the Anderson-Darling test, the decision criterion is to reject the null hypothesis when the test statistic is greater than or equal to the 5\% critical value. 

The Shapiro-Wilk test examines the null hypothesis that a sample of data is drawn from a normal distribution~\citep{shapiro_analysis_1965}.  The Shapiro-Wilk test was implemented via the \texttt{shapiro} function in the \texttt{SciPy} library, which provides both the test statistic and the corresponding $p$-value~\citep{guthrie_nistsematech_2020}. The Shapiro-Wilk test a considered a very reliable test, especially for small to medium sample sizes (50 < n < 2000). 

The D'Agostino's $K^2$ test examines normality by assessing whether the skewness and kurtosis of the sample distribution differ significantly from those expected under normality~\citep{dagostino_omnibus_1971}. The D'Agostino's $K^2$ test for normality was performed using the \texttt{normaltest} function from the \texttt{SciPy} library, which returns both the test statistic and the corresponding $p$-value~\citep{guthrie_nistsematech_2020}.

The final normality test examined is the Anderson-Darling test, a modification of the Kolmogorov-Smirnov test. The Anderson-Darling test evaluates the distribution of the sample data by comparing its empirical distribution to a theoretical distribution~\citep{anderson_test_1954, kolmogorov_sulla_1933, smirnov_table_1948}. The Anderson--Darling test for normality was conducted using the \texttt{anderson} function from the \texttt{SciPy} library, which provides the test statistic and the corresponding critical value at the 5\% level of significance~\citep{guthrie_nistsematech_2020}.

\subsection{Test of Hypothesis: Variance Difference between Historical and Forecasted Demand}\label{subsec:mt_variance_test}

To address the limitation of backward-looking safety stock dimensioning, a test of difference in variation between historical demand and forecasted demand was conducted. The specific statistical formulation examined follows \citet{brown_robust_1974}. The null hypothesis investigated was that there is no significant difference in variance, whereas the alternative hypothesis was that there is a significant difference in variance. To compare the variability of historical consumption and forecasted demand, we employ the robust Levene/Brown–Forsythe (BF) test, which assesses equality of variances while being less sensitive to non-normality. Specifically, we use the Brown–Forsythe variant that centers absolute deviations around the sample median~\citep{brown_robust_1974}. Let $y_{gj}$ denote observation $j$ in group $g\in\{1,2\}$ (historical consumption vs.\ forecasted demand), and let $\tilde{y}_g$ be the group median. Define $z_{gj}=\lvert y_{gj}-\tilde{y}_g\rvert$. The test statistic is the one-way ANOVA $F$ statistic computed on $\{z_{gj}\}$:
\[
F=\frac{(N-k)\sum_{g=1}^{k} n_g(\bar{z}_g-\bar{z})^2}{(k-1)\sum_{g=1}^{k}\sum_{j=1}^{n_g}(z_{gj}-\bar{z}_g)^2},\quad k=2,
\]
which under $H_0:\sigma_1^2=\sigma_2^2$ approximately follows an $F_{k-1,\,N-k}$ distribution. We implement this via \texttt{scipy.stats.levene(..., center='median')}. Reported $p$-values and decisions in Table~\ref{tab:variance} are based on this BF test. When the p-value of the F-test statistic is less than the 5\% significance level, the null hypothesis is rejected, indicating that the variability between historical and forecasted demand differs. This finding suggests that relying solely on historical variance may not capture the full spectrum of uncertainty influencing future demand conditions. Accordingly, the following section builds on this result by pooling both historical and forecasted demand data to form a joint empirical distribution that reflects their combined variability.

\subsection{Forward-Looking Variability Estimation}\label{subsec:mt_forward_var}

The leading approach in the empirical literature for addressing the limitations of demand variability is to include forecast-error variability. Forecast error variability is not without its challenges. Particularly, forecast error variability does not incorporate forward-looking demand signals for safety stock dimensioning. This study proposes a forward-looking demand-variability estimation by combining historical and forecasted demand. This approach does not assume equal variances across sources; instead, it leverages their joint variability to better capture potential demand dynamics. The formal specification used is stated in (4).
\begin{equation}
    D^{*} = D_{c} \,\Vert\, D_{f},
    \label{eq:eq_trad_4}
\end{equation}
such that, ``$\Vert$'' denotes concatenation of the historical demand and forecasted demand, $n = |D_{c}| + |D_{f}|$, $ \mu = \frac{1}{n}\sum_{i=1}^{n} D_i^* $, $ \sigma^2 = \frac{1}{n-1} \sum_{i=1}^{n} (D_i^* - \mu)^2 $,
where $D^*$ = combined demand, $D_c$ = historical consumption, $D_f$ = forecasted demand, $\mu$ = combined mean and, $\sigma^2$ = forward-looking variance.

\subsection{Nonparametric Distribution Estimation}\label{subsec:mt_dist_estimation}
Empirically, assuming that demand is normally distributed is not valid for many reasons. First, in the real world, demand exhibits skewness and kurtosis, violating the classical assumptions of normality. Second, there are many circumstances in which demand is discretely distributed. Lastly, assuming the same demand distribution for each item is not practical. To relax the normality assumption and approximate the custom per-item demand distribution, Kernel Density Estimation (KDE), a nonparametric distribution-estimation method, was implemented. KDE provides a data-driven approximation of the probability density function without assuming any parametric form~\citep{parzen_estimation_1962,plesovskaya_empirical_2021,kolmogorov_sulla_1933}. For implementation details as specified in (\ref{eq:eq_trad_5}), the Gaussian kernel was preferred for its analytical tractability and consistency properties under Scott's bandwidth rule \citet{scott_multivariate_2015}, balancing bias and variance as recommended by \citet{silverman_density_1998}.
\begin{equation}
    \hat{f}(x) = \frac{1}{nh} \sum_{i=1}^{n} K\left( \frac{x-D_i^*}{h} \right)
    \label{eq:eq_trad_5}
\end{equation}
where $\hat{f}(x)$ = the estimated probability density function at point $x$, $K(.)$ = Gaussian kernel function, specified as $K(u)=\frac{1}{\sqrt{2\pi }}e^{-\frac{1}{2}u^{2}}$, $h$ = bandwidth (in this case Scott’s rule: $h=n^{-1/5}\sigma$), $n$ = number of data points, $D_i^*$ = individual demand data point.

Following the procedures outlined by \citet{aitchison_multivariate_1976}, \citet{wand_kernel_1994}, and \citet{chen_probability_2000}, the continuous KDE was transformed into a probability mass function (PMF). Specifically, we evaluated the density values at integer support points obtained from the combined demand sample, $D^*$,  and normalized such that $P(X = x_j) = \frac{\hat{f}(x_j)}{\sum_j \hat{f}(x_j)}$, $\sum_j P(X = x_j) = 1$. The discretization results in a valid nonparametric distribution that preserves the smoothed KDE structure while adhering to non-negative discrete demand realizations. It is worth noting that the implementation details specified in the accompanying codebase, corrected for boundary issues by reassigning negligible negative probabilities to the lowest demand support point, thus ensuring non-negativity and the conservation of total probability mass.

Empirically, KDE has been shown to outperform normal and Poisson assumptions in modelling demand uncertainty in manufacturing contexts~\citep{eaves_forecasting_2004, hasni_investigation_2019}. It accurately represents multimodal, skewed demand and yields expected service levels. Recent studies also confirm the robustness of KDE under small to moderate samples \citet{plesovskaya_empirical_2021} or in volatile supply chains \citet{syntetos_distributional_2011}. The resulting discrete nonparametric PMF serves as input to the stochastic simulation described in Section~\ref{subsec:mt_simulation}, enabling data-driven safety stock estimation that remains valid even when classical distributional assumptions are violated.

\subsection{Simulation Model}\label{subsec:mt_simulation}
Monte Carlo simulation is a robust method for analyzing stochastic inventory systems~\citep{axsater_inventory_2006,snyder_fundamentals_2019,sunil_chopra_supply_2025}. To evaluate the performance of the safety stock computed under both the Normal and Kernel Density Estimation (KDE) achieves the desired service levels, a stochastic simulation model was developed. Specifically, the demand distributions obtained in Section~\ref{subsec:mt_dist_estimation} were used as inputs to a simulation model that replicates the real-world dynamics of inventory replenishment: random demand, replenishment lead times, and review cycles. The approach aligns with empirical studies by \citet{musalem_controlling_2005} and \citet{chu_simulation-based_2014}, which showed that simulation provides a more accurate reflection of system performance than analytical approximations, particularly when demand is non-normal.

\subsubsection*{Simulation Design}

For each item, the simulation model takes in as input the simulation period $T$, the review period $R$, and the lead time $L$ (all in weeks). Also, the simulation takes the fitted KDE-based probability mass function from Section~\ref{subsec:mt_dist_estimation} as input for the KDE-based model, or the parameters of a normal distribution $N(\mu, \sigma^2)$ as input for the normal-based model. The simulation operates over $T$ periods and models a periodic-review $(R,S)$ inventory policy. Before the simulation begins, random demand array $D_T$ of length $T$ is generated from either the fitted KDE-based probability mass function or from a normal distribution $N(\mu, \sigma^2)$. Demand at each period t is denoted as $D_t$. Then the safety stock $S_s$ and the base-stock level $S$ are determined for each model. The safety stock is defined as follows:
\begin{equation}
    S_s = \begin{cases}
    Q_\alpha(D_{L+R}) - \mathbb{E}[D_{L+R}], & \text{KDE model} \\
   {Z_\alpha\sigma\sqrt{L+R}}, & \text{Normal model}
    \end{cases}
    \label{eq:eq_trad_6}
\end{equation}
where $Z_\alpha$ is the standard normal quantile for service level $\alpha$ and $Q_\alpha(D_{L+R})$ denotes the $\alpha$-quantile of total demand over the risk horizon $(L+R)$, obtained from $T$  periods of Monte Carlo samples.

The base stock is computed as:
\begin{equation}
    S = S_s + 2C_s + I_s
    \label{eq:eq_trad_7}
\end{equation}
where cycle stock (stock determined for normal demand) $C_s =  \frac{1}{2}\mu R$ and in-transit inventory (inventory ordered from supplier but not yet available in warehouse) $I_s= \mu L$. 

At each period $t$, demand $D_t$ is obtained from the random demand array $D_T$. The on-hand inventory, $H_t$, is inventory on-hand at the beginning of each simulation time step. Transit inventory is modelled explicitly as a pipeline vector $T_{t,k}$ for $k=0,1,…,L-1$, representing the number of periods remaining until order arrival. At each period, items advance by one period toward receipt such that:

\begin{equation}
    T_{t,k}= T_{t-1,k+1}
    \label{eq:eq_trad_8}
\end{equation}
The on-hand inventory, $H_t$, is updated according to the inventory balance equation:

\begin{equation}
    H_t = \max(0, H_{t-1} - D_t + T_{t-1,0})
    \label{eq:eq_trad_9}
\end{equation}
where $H_{t-1}$ is the previous on-hand stock and $T_{t-1,0}$ denotes items whose lead time just expired (arriving this period). The simulation assumes a lost-sales condition—demand exceeding available inventory is lost rather than backordered—to reflect typical manufacturing replenishment behaviour. This structure mirrors the stochastic inventory dynamics formalized in \citet{silver_biased_1987} and \citet{axsater_inventory_2006}. A replenishment order is triggered every Review, $R$, period according to the policy:

\begin{equation}
   T_{t,L} = S - \left( H_t + \sum_{k=0}^{L-1} T_{t,k} \right)
    \label{eq:eq_trad_10}
\end{equation}
such that the inventory position (on-hand inventory plus in-transit inventory) is replenished to the base-stock level $S$. This replenishment mechanism is not different from the classical review system formulations in \citet{zipkin_foundations_2000} and \citet{stadtler_purchasing_2015}.

\subsubsection*{Performance Measurement}

The simulation outputs three service-level metrics:

\begin{enumerate}
\renewcommand{\labelenumi}{(\roman{enumi})}
    \item \textbf{Cycle service level}: the probability of completing a replenishment cycle without a stockout. The cycle service level is calculated as:
    
        \begin{equation}
            SL_{\text{cycle}} = 1 - \frac{\sum_{\text{cycles}} \mathbb{I}{\text{[stockout in cycle]}}}{\# \text{ of cycles}}
            \label{eq:eq_trad_11}
        \end{equation}
        
    \item \textbf{Period service level}: the proportion of total periods without stockout. The period service level is calculated as:
    
        \begin{equation}
            SL_{\text{period}} = 1 - \frac{\sum_t SO_t}{T}
            \label{eq:eq_trad_12}
        \end{equation}
    where $SO_t$ = Stock out at period $t$. Binary; 1 if $H_t$ = 0, 0 if $H_t$ > 0. 
    
    \item \textbf{Safety stock value}: the total value of the safety stock, computed as:
        \begin{equation}
            SS_{\text{value}} = S_s \times PU
            \label{eq:eq_trad_13}
        \end{equation}
    where $PU$ is the per-unit cost of an item.
\end{enumerate}

The simulation was executed for each item across service-level targets ranging from 0\% to 99\%. For each configuration, we simulated 1000 periods to ensure convergence of empirical service levels. The resulting safety stock, realized service levels, and total safety stock costs were recorded as outputs for subsequent optimization in Section~\ref{subsec:mt_linear_prog} to identify cost-minimizing service-level configurations across items. A summary of the simulation algorithm is presented in Algorithm~\ref{alg:simulation}. Refer to the \href{https://github.com/graphshade/safety_stock_simulation}{GitHub} code base for the actual Python implementation.



\begin{algorithm}
    \caption{Simulation Algorithm}
    \label{alg:simulation}
    \begin{algorithmic}[1] 
        \Require $L, R, P(X = X_j)$
        \Ensure $SL_{\text{cycle}}$, $SL_{\text{period}}$ and $SS_{\text{value}}$
        \State Compute $S_s, C_s, I_s, S$
        \State Initialize arrays: $H_T \leftarrow  \mathbf{0_T}$, $SO_t \leftarrow  \mathbf{0_T}$,  $ T_{t,L} \leftarrow \mathbf{0}_{t\times K}$
        \State Set $H_0 = S - D_0$, $T_{0,L} = D_0$
        \For{$t \gets 1$ to $T \gets 1000$}
             \State Get demand $D_t$ at time $t$
            \State Update $H_t = \max(0, H_{t-1} - D_t + T_{t-1,0})$
            \State Shift in-transit inventory pipeline: $T_{t,k} = T_{t-1,k+1}$
            \State Place a new order if $t \bmod R = 0$. Order size, $T_{r,L} = S - \left(H_t + \sum_{k=0}^{L-1} T_{t,k}\right)$
            \State Record stockout indicator $SO_t$
        \EndFor
        \State Compute $SL_{\text{cycle}}$, $SL_{\text{period}}$, and $SS_{\text{value}}$
        \State \Return $SL_{\text{cycle}}$, $SL_{\text{period}}$ and $SS_{\text{value}}$
    \end{algorithmic}
\end{algorithm}

\subsection{Linear Programming Optimization}\label{subsec:mt_linear_prog}
The final aspect of the proposed framework integrates the simulation outputs into a prescriptive linear programming (LP) model to identify the optimal service-level configuration across the items. The LP model minimizes total safety stock value while satisfying predetermined category-wise service-level constraints, aligning operational priorities with cost efficiency in inventory control. The approach adopted builds on foundational operations research principles \citet{dantzig_linear_1965, hillier_introduction_2021, winston_operations_2004} and on the application of optimization in supply chain management \citet{axsater_inventory_2006, silver_biased_1987}.

\subsubsection*{Model Formulation}

Let $i$ represent the item, $k$ represent the potential service levels (e.g., 0.50, 0.60, 0.70, 0.80, 0.90, 0.95, 0.99), and c represent the item class priority (e.g., A, B, C). The binary decision variable $x_{ik}$  equals 1 if item $i$ adopts service level $k$, and 0 otherwise. The optimization problem is formulated as:

\begin{center}
$\min_{X_{ik}} \sum_{i,k} C_{ik} X_{ik}$
\end{center}

\textbf{subject to:} \\

\begin{align*}
& \sum_{k} X_{ik} = 1, \quad \forall i \\
& 
\forall\, c \in \mathcal{C}:\quad 
\sum_{i \in \mathcal{I}_c} \sum_{k} W_{ic}\,\alpha_{ik}\,X_{ik} \;\ge\; WSL_c^{\text{target}} \\
& X_{i,k} \in \{0,1\} 
\end{align*}

\noindent
\textbf{where:}
\begin{itemize}
    \item $C_{ik}$ is the total safety stock cost or value for item $i$ at service level $k$;
  \item $\mathcal{C}$ is the set of all product categories (or classes);
  \item $\mathcal{I}_c$ is the subset of items belonging to category $c$;
  \item $W_{ic}$ is the weight associated with item $i$ in category $c$;
  \item $\alpha_{ik}$ is the $SL_{cycle}$ corresponding to item  $i$ at service level $k$;

  \item $WSL_c^{\text{target}}$ is the target weighted service level for category $c$.
\end{itemize}

\noindent



The coefficients $C_{ik}$ and $\alpha_{ik}$ are obtained from the simulation described in Section~\ref{subsec:mt_simulation}. For each combination of item $i$ and service level $k$, the simulation provides the realized service level performance ($SL_{\text{cycle}}$, $SL_{\text{period}}$), the required safety stock ($S_s$), and the associated safety stock value ($SS_{\text{value}}$). These parameters are then fed into the LP model. The simulation-optimization approach follows the data-driven approach implemented by \cite{fu_feature_2002} and later applied empirically in supply chain contexts~\citep{chu_simulation-based_2014, zotteri_impact_2005}) where simulation-derived parameters serve as deterministic inputs for optimization.

The LP problem is solved using a branch-and-bound algorithm via the PulP CBC solver (see the \href{https://github.com/graphshade/safety_stock_simulation}{GitHub} code base for the Python implementation). The solver identifies the service-level configuration that minimizes total safety stock cost while maintaining weighted service levels appropriate to item categories and priority classes (A/B/C and special items). In the specific implementation for the 20 items studied, the $WSL_c^{\text{target}}$ for class A and B items were 95\% and 90\% respectively. The classification-based constraint approach follows the prioritization framework of \citet{eaves_forecasting_2004} and \citet{tempelmeier_inventory_2006}, which recommends differentiated service-level targets across item categories.

\section{Results}\label{sec:results}

\subsection{Normality Test Results}\label{subsec:results_nor_test}
\begin{figure}[h]
  \centering
  \includegraphics[width=0.8\textwidth]{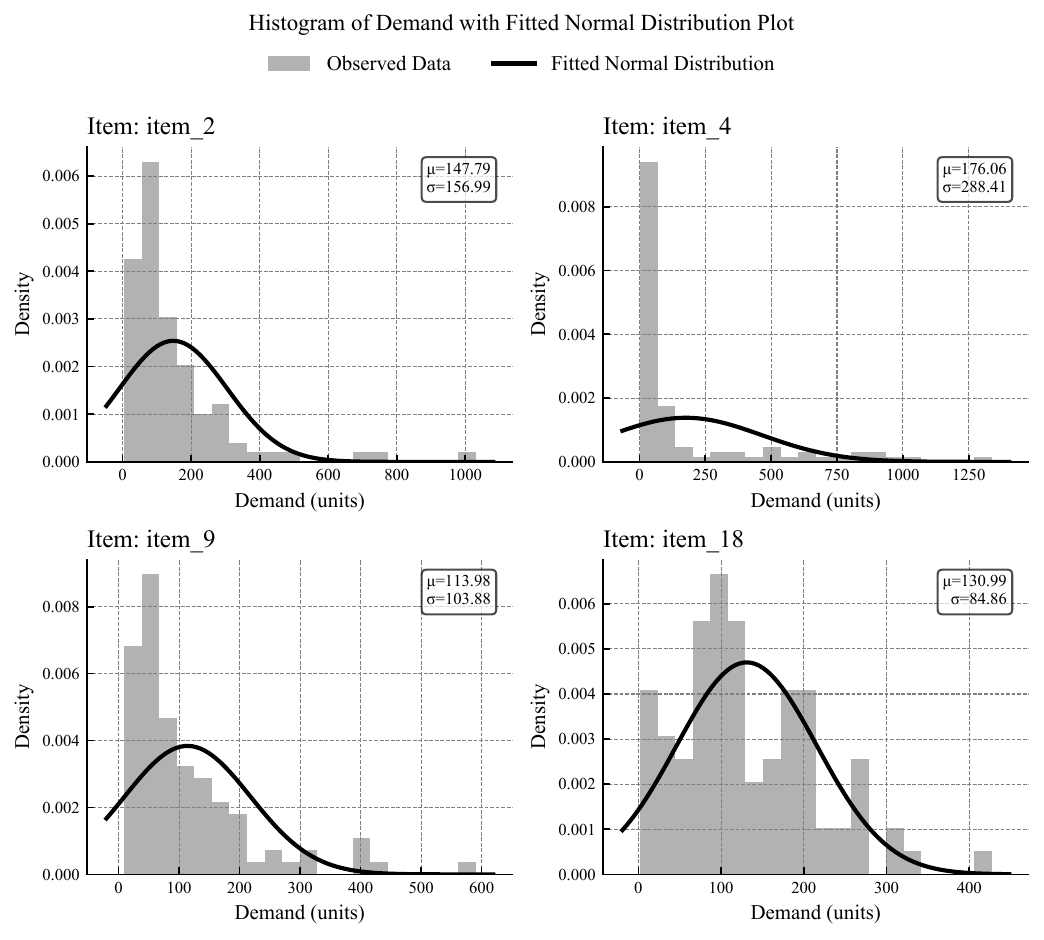}
  \caption{Density distribution of demand (in units) for items 2, 4, 9, and 18: Shows the frequency of demand levels for each item, with a density curve superimposed on each histogram to illustrate the distribution's shape.
  }
  \label{fig:histogram}
\end{figure}

Figure~\ref{fig:histogram} is a histogram of four randomly selected items. Visually, the distributions of items 2, 4, and 9 exhibit greater right skewness than that of item 18, suggesting that the demand distributions for the selected items are not normally distributed. The normality of the demand distributions for all items was evaluated using the Shapiro-Wilk, D'Agostino's $K^2$, and Anderson-Darling tests. The test statistics, p-values, and critical values (at the 5\% level of significance) for the corresponding tests are reported in Table~\ref{tab:normality}. Based on the results presented in Table~\ref{tab:normality}, the normality assumption in traditional safety stock diminishing techniques does not hold.

\begin{table}[ht]
\centering
\begin{tabular}{lrrrrrr}
\toprule
item & t-stat\_SH & t-stat\_DA & t-stat\_AN & c-val\_AN & p-val\_SH & p-val\_DA \\
\midrule
item\_1 & 0.75 & 63.43 & 6.51 & 0.76 & 0.00 & 0.00 \\
item\_2 & 0.69 & 87.58 & 7.40 & 0.76 & 0.00 & 0.00 \\
item\_3 & 0.65 & 68.93 & 11.16 & 0.76 & 0.00 & 0.00 \\
item\_4 & 0.64 & 47.84 & 14.27 & 0.76 & 0.00 & 0.00 \\
item\_5 & 0.76 & 64.77 & 6.19 & 0.76 & 0.00 & 0.00 \\
item\_6 & 0.84 & 38.27 & 3.88 & 0.76 & 0.00 & 0.00 \\
item\_7 & 0.62 & 41.84 & 10.15 & 0.75 & 0.00 & 0.00 \\
item\_8 & 0.95 & 8.49 & 1.07 & 0.76 & 0.00 & 0.01 \\
item\_9 & 0.78 & 54.17 & 6.14 & 0.76 & 0.00 & 0.00 \\
item\_10 & 0.90 & 21.52 & 2.79 & 0.76 & 0.00 & 0.00 \\
item\_11 & 0.81 & 30.77 & 6.44 & 0.76 & 0.00 & 0.00 \\
item\_12 & 0.89 & 32.34 & 2.49 & 0.76 & 0.00 & 0.00 \\
item\_13 & 0.90 & 37.60 & 1.78 & 0.76 & 0.00 & 0.00 \\
item\_14 & 0.38 & 133.85 & 19.10 & 0.76 & 0.00 & 0.00 \\
item\_15 & 0.86 & 36.99 & 3.52 & 0.76 & 0.00 & 0.00 \\
item\_16 & 0.88 & 19.38 & 3.85 & 0.76 & 0.00 & 0.00 \\
item\_17 & 0.82 & 56.31 & 3.58 & 0.76 & 0.00 & 0.00 \\
item\_18 & 0.95 & 10.05 & 0.98 & 0.76 & 0.00 & 0.01 \\
item\_19 & 0.81 & 59.97 & 4.02 & 0.76 & 0.00 & 0.00 \\
item\_20 & 0.87 & 27.34 & 3.70 & 0.76 & 0.00 & 0.00 \\
\bottomrule
\end{tabular}
\caption{Results of Shapiro-Wilk(SH), D'Agostino(DA), and Anderson-Darling(AN) Normality Tests: the table shows test-statistics (t-stat), critical-value (c-val) and p-values (p-val) for the various tests. At the 5\% significance level, there is statistically significant evidence that the demand for each item is not normally distributed.} 
\label{tab:normality}
\end{table}

\subsection{Comparison of Variability Estimates}\label{subsec:results_var_test}
Safety stock dimensioning based on forecast error variability alone considers only historical variability being computed from past data that may not fully reflect future demand volatility. To evaluate this assertion, the hypothesis that there is a significant difference in variance between historical and forecasted demand was examined using Levene's test for equality of variances \citet{brown_robust_1974}. Table ~\ref{tab:variance} presents the results of Levene's test. As the p-value for each item is less than the 5\% significance level, there is evidence that historical demand variance differs significantly from forecasted demand variance.

\begin{table}[!ht]
\centering
\begin{tabular}{lrr}
\toprule
item & levene\_f-statistics & p-value \\
\midrule
item\_1 & 18.98 & 0.00 \\
item\_2 & 20.66 & 0.00 \\
item\_3 & 22.09 & 0.00 \\
item\_4 & 15.35 & 0.00 \\
item\_5 & 21.53 & 0.00 \\
item\_6 & 22.93 & 0.00 \\
item\_7 & 21.05 & 0.00 \\
item\_8 & 31.46 & 0.00 \\
item\_9 & 22.55 & 0.00 \\
item\_10 & 26.68 & 0.00 \\
item\_11 & 20.04 & 0.00 \\
item\_12 & 31.86 & 0.00 \\
item\_13 & 20.06 & 0.00 \\
item\_14 & 5.95 & 0.02 \\
item\_15 & 30.08 & 0.00 \\
item\_16 & 25.97 & 0.00 \\
item\_17 & 31.65 & 0.00 \\
item\_18 & 15.18 & 0.00 \\
item\_19 & 15.47 & 0.00 \\
item\_20 & 18.91 & 0.00 \\
\bottomrule
\end{tabular}

\caption{Levene's Test for Homogeneity of Variances: shows the f-statistics and the p-value for the Levene's test. At the 5\% level of significance, there exists evidence that historical demand variance is statistically different from forecasted demand variance for each item.} 
\label{tab:variance}
\end{table}

\subsection{Simulation Performance}\label{subsec:results_sim_results}
\begin{figure}[h!]
  \centering
  \includegraphics[width=\textwidth, height=1.25\textwidth, keepaspectratio]{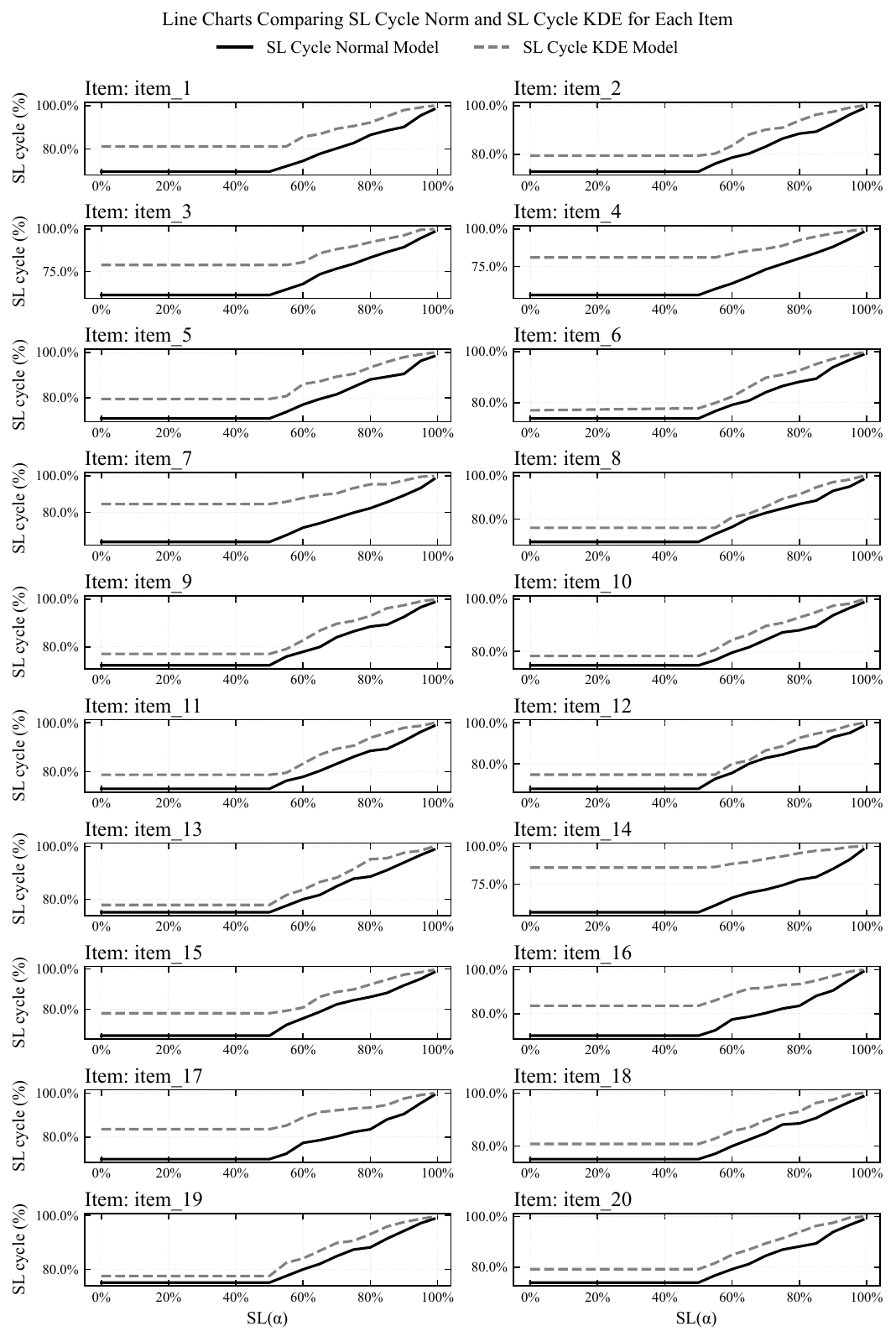}
  \caption{Comparison of realized cycle service levels from two safety stock simulation models: The plot shows the realized service level ($SL_{\text{cycle}}$) for each of the two models, one based on a normal distribution and the other on Kernel Density Estimation (KDE), plotted against the target service level input ($SL(\alpha)$) for twenty different items.
  }
  \label{fig:simulation}
\end{figure}

Figure~\ref{fig:simulation} reports the results of the simulation model specified in Section~\ref{subsec:mt_simulation}. Figure~\ref{fig:simulation} is a line plot of the realized service level ($SL_{\text{cycle}}$) for each of the two models---the normal model and the KDE-based model---plotted against the target service level input ($SL(\alpha)$) for twenty different items. The graph shows that dimensioning safety stock using a nonparametric approach, such as KDE, is more robust and accurate than the traditional approach of assuming normality. Specifically, for a given service level input (e.g., 80\%) the realized service level is higher under the KDE-based model than the normal model. The normality assumption for demand distribution leads to a systematic underestimation of the safety stock needed to meet the target service level, resulting in a lower realized service level. The KDE-based model, by contrast, does not assume a specific underlying probability distribution like the normal model does. It approximates each item's actual demand patterns, resulting in a much better match between the desired and realized service levels.

\subsection{Optimization Results }\label{subsec:results_opt_results}
We evaluated the optimization model specified in Section~\ref{subsec:mt_linear_prog} for the proposed forward-looking KDE-based approach and the traditional normal distribution approach. Table~\ref{tab:optimization} summarizes the results of the optimization model, showing the expected cycle service level, the expected period service level and the safety stock value by item class. As shown in Table~\ref{tab:optimization}, the optimization results for the KDE-based model are not only cost-efficient but also expected to achieve comparatively higher cycle and period service levels than the normal model across all item classes. For instance, for the higher-priority items, the forward-looking KDE-based approach is expected to achieve 94\% and 97\% cycle service and period service levels, respectively, using \$1.8M less safety stock than the normal model. It is important to note that the optimization constraint $\sum_{i\in\mathcal{I}_c}\sum_k W_{ic}\,\alpha_{ik}\,X_{ik}\ge WSL_c^{\text{target}}$ is formulated in terms of a \emph{weighted} service level, where $W_{ic}$ represents the relative weight of item~$i$ in class~$c$. Accordingly, the “Expected Cycle SL’’ values reported in Table~\ref{tab:optimization} are unweighted averages across items within each class and are presented for interpretability. While these simple averages may fall slightly below the class-level target, the weighted service levels used in the optimization satisfy the required thresholds of 95\% and 90\%, respectively. Thus, the linear-programming solutions remain feasible with respect to the model’s service-level constraints. 

\begin{figure}[h!]
  \centering
  \includegraphics[width=\textwidth]{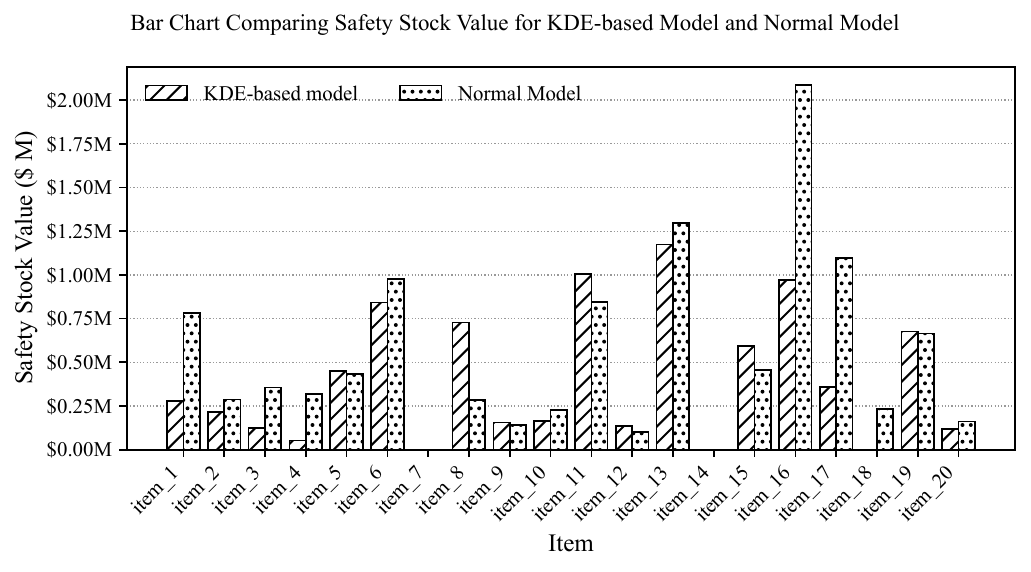}
  \caption{Item-wise comparison of safety stock values for the KDE-based model and the normal model: Final safety stock value from the optimization model using the KDE-based model and the normal model.
  }
  \label{fig:item-wise}
\end{figure}

The item-wise bar chart in Figure~\ref{fig:item-wise} provides a breakdown of the total safety stock value, as determined by the optimization model. The total safety stock value is lower for the KDE-based model than for the normal model for all items except item\_8 and item\_11.

\begin{table}[ht]
\centering
\begin{tabular}{llcccc}
\toprule
Item Class & Model & Expected Cycle SL & Expected Period SL & Safety Stock Value & \\
\midrule
\multirow{2}{*}{A} & KDE-based Model & 93.69\% & 96.93\% & \textdollar 6.63M & \\
& Normal Model & 91.65\% & 95.83\% & \textdollar 8.45M & \\
\cmidrule(lr){2-6}
\multirow{2}{*}{B} & KDE-based Model & 89.3\% & 94.77\% & \textdollar 1.41M & \\
& Normal Model & 88.54\% & 94.49\% & \textdollar2.26M & \\
\bottomrule
\end{tabular}
\caption{Summary of optimization results: optimization results, including the expected cycle and period service levels, and the total safety stock value for the KDE-based model and the Normal model.} 
\label{tab:optimization}
\end{table}

\section{Discussion and Managerial Implication}\label{sec:discussion}
Safety stock is a crucial inventory control mechanism for managing demand and supply uncertainty. The quality of the dimensioned safety stock is even more critical because it affects inventory holding costs and end-customer service levels. For instance, a high safety stock setting guarantees a higher level of supply chain serviceability, but it comes at a higher inventory holding cost. In contrast, a lower safety stock will keep inventory holding costs low but will result in comparatively poor supply chain serviceability. We propose a data-driven analytical approach to dimension safety stock to ensure the computed safety stock achieves the desired serviceability at an optimal inventory level. A Monte Carlo simulation with 1000 periods was used to evaluate the performance of the proposed approach compared to the traditional approach for dimensioning safety stock, assuming normally distributed demand. The simulation model outputs are passed to a linear optimization model to determine the optimal safety stock configuration.

The comparative summary results in Table~\ref{tab:optimization} show that the proposed analytical approach achieves higher cycle and period service levels at a lower inventory value than the traditional approach. Evidently, using a nonparametric approach, where no assumption is made about the demand distribution for each item, resulted in a KDE-based probability mass function that is more robust at capturing item-specific demand patterns than traditional approaches that assume normality. Another important observation is that incorporating forecasted demand variability aligns with the fundamental notion underpinning safety stock: setting a buffer stock to protect against unexpected demand beyond the forecast. 

The data-driven analytical approach adopted in this paper ensures robust study results and high business relevance. The first analytical design was to use KDE, a nonparametric approach, to estimate the demand distribution and convert the resulting continuous distribution to a probability mass function. This analytical approach to estimating the demand distribution ensures that each item's demand distribution is modelled independently. Also, converting from a continuous distribution to a probability mass function is business-relevant because demand in many business contexts is discrete and positive. The simulation results presented in Figure~\ref{fig:simulation} show that using KDE provides a robust estimate of the demand pattern for each item compared to a parametric approach such as a normal distribution. 

The second analytical technique adopted is the inclusion of forecasted demand in the estimation of demand variability. The intent is to capture forward-looking variability rather than relying solely on forecast-error variability. The results in Table~\ref{tab:optimization} show that including forward-looking variability ensures that the safety stock computed is not overestimated based on historical variability. 

Lastly, we used a simulation approach to evaluate the results of the proposed safety stock modelling approach and the traditional method. The results of the simulation design were passed to a linear programming model to determine the optimal safety stock configuration. The simulation technique ensures evaluation of the models under near-reality scenarios, and the linear programming model ensures business relevance of attaining desired service levels under specific business constraints. 

The results of the proposed data-driven approach to safety stock modelling are more robust than those of the traditional approach. Chiefly, the use of KDE for the probability mass function performs better at modelling individual-item demand patterns. Also, incorporating forecast demand variability improves the final safety stock to be forward-looking and less prone to overestimation or underestimation compared with approaches based solely on historical variability. While the proposed data-driven method is robust, it assumes that the demand distribution for each item is independent, which may be restrictive in supply chains with modular product design, where the demand distribution of one item depends on that of others. Another limitation of the proposed framework is that including forward variability means the quality of the dimensioned safety stock depends on the forecast quality. Given the limitations mentioned previously, future work should explore multivariate nonparametric approaches that can capture the underlying interdependent item-complex distributions. Future research should also evaluate the sensitivity of the proposed analytical approach to forecast uncertainty by using forecast-error variance as a diagnostic metric. Such analyses would clarify how variations in forecast quality influence the robustness and reliability of the KDE-based dimensioning results.

\section{Conclusion}\label{sec:conclusion}
Safety stock management is paramount to achieving resilient supply chains for hardware OEMs. However, the prevalent methods for dimensioning safety stock fail in approximating the demand conditions of inventory components. To address this issue, we propose a data-driven analytical approach in which no prior assumptions are made about the demand distribution, and forecasted demand variation is included in estimating overall demand variability. A Monte Carlo simulation of 1000 periods was used to evaluate the performance of the proposed approach vis-à-vis the traditional approach of dimensioning based on demand variance, which assumes normality of the demand distribution for all individual items. The simulation results were used as inputs into a linear optimization model that determines the optimal safety stock setting that satisfies business-specific constraints, ensuring optimal inventory cost. 
The results obtained using the analytical approach showed robust performance compared to the classical approach to safety stock dimensioning. Notably, the proposed approach achieved higher cycle and period service levels at a lower inventory value than the traditional approach. For instance, for high-priority items, the cycle and period service levels achieved using the proposed approach were 2 and 1 percentage points higher than the traditional approach, respectively, but at a \$1.8 million lower inventory value. 

Notwithstanding the robust performance of the proposed KDE-based approach, the study assumed that item-wise demand distributions are independent, which may be limiting in a modular manufacturing context characterized by a high item commonality index. As a result, future studies should explore the use of multivariate KDE to model the demand distribution.

\newpage

\printbibliography

\end{document}